\documentclass[aps,prl,twocolumn,showpacs,superscriptaddress,floatfix, nofootinbib, nobibnotes]{revtex4-2}
\usepackage[caption=false]{subfig}
\usepackage{amsmath,graphicx}
\usepackage{amsmath,amssymb,mathrsfs,esint, tikz}
\usepackage{textcomp}
\usepackage{multirow}
\usepackage{float}
\makeatletter
\let\newfloat\newfloat@ltx
\makeatother
\usepackage{algorithm}
\usepackage[noend]{algpseudocode}
\usepackage{xpatch}
\usepackage{comment}
\usepackage{tabularx, braket}
\usepackage{bm}
\usepackage[normalem]{ulem}
\usepackage[bookmarks=true,
   colorlinks=true,
   linkcolor=blue,
   urlcolor=blue,
   citecolor=blue,
   bookmarks=true,
   hyperindex=true
]{hyperref}
\usetikzlibrary{quantikz2}

\newcommand{\cnx}{\ensuremath{{{\sf C}^{N-1}\sf{X}}}}

\makeatletter
\xpatchcmd{\algorithmic}{\itemsep\z@}{\itemsep=1ex plus2pt}{}{}
\makeatother

\tikzset{U1/.style={
    minimum width=1cm,
    minimum height=1cm,
    path picture={
        \begin{scope}[x=2cm, y=0.5cm]
        \draw[->, violet!40, ultra thick] (path picture bounding box.center) +(-0.19,-4) -- +(-0.19,4);
        \draw[->, cyan!40, ultra thick] (path picture bounding box.center) +(0.19,-4) -- +(0.19,4);
        \end{scope}
    }
}}

\tikzset{U1last/.style={
    minimum width=1cm,
    minimum height=1cm,
    path picture={
        \begin{scope}[x=2cm, y=0.5cm]
        \draw[->, violet!40, ultra thick] (path picture bounding box.center) +(-0.19,-4) -- +(-0.19,4);
        \end{scope}
    }
}}

\tikzset{U1lastdag/.style={
    minimum width=1cm,
    minimum height=1cm,
    path picture={
        \begin{scope}[x=2cm, y=0.5cm]
        \draw[->, cyan!40, ultra thick] (path picture bounding box.center) +(0.19,-4) -- +(0.19,4);
        \end{scope}
    }
}}

\tikzset{U2/.style={
    minimum width=1cm,
    minimum height=1cm,
    path picture={
        \begin{scope}[x=2cm, y=0.5cm]
        \draw[->, cyan!20, ultra thick] (path picture bounding box.center) +(-0.19,-4) -- +(-0.19,4);
        \draw[->, violet!40, ultra thick] (path picture bounding box.center) +(0.19,-4) -- +(0.19,4);
        \end{scope}
    }
}}

\definecolor{newgreen}{RGB}{77, 175, 74}
\definecolor{nblue}{RGB}{55, 126, 184}

\newcommand{\RB}[1]{{\color{black} #1}}
\newcommand{\RC}[1]{{\color{black} #1}}

\begin{document}

\title{Scalable improvement of the generalized Toffoli gate realization \\ using trapped-ion-based qutrits}

\author{Anastasiia S. Nikolaeva}
\thanks{These two authors contributed equally.
Correspondence should be
addressed to: anastasiia.nikolaeva21@gmail.com}
\affiliation{P.N. Lebedev Physical Institute of the Russian Academy of Sciences, Moscow 119991, Russia}
\affiliation{Russian Quantum Center, Skolkovo, Moscow 121205, Russia}

\author{Ilia V. Zalivako}
\thanks{These two authors contributed equally.
Correspondence should be
addressed to: anastasiia.nikolaeva21@gmail.com}
\affiliation{P.N. Lebedev Physical Institute of the Russian Academy of Sciences, Moscow 119991, Russia}
\affiliation{Russian Quantum Center, Skolkovo, Moscow 121205, Russia}

\author{Alexander S. Borisenko}
\affiliation{P.N. Lebedev Physical Institute of the Russian Academy of Sciences, Moscow 119991, Russia}
\affiliation{Russian Quantum Center, Skolkovo, Moscow 121205, Russia}

\author{Nikita V. Semenin}
\affiliation{P.N. Lebedev Physical Institute of the Russian Academy of Sciences, Moscow 119991, Russia}
\affiliation{Russian Quantum Center, Skolkovo, Moscow 121205, Russia}

\author{Kristina P. Galstyan}
\affiliation{P.N. Lebedev Physical Institute of the Russian Academy of Sciences, Moscow 119991, Russia}
\affiliation{Russian Quantum Center, Skolkovo, Moscow 121205, Russia}

\author{Andrey E. Korolkov}
\affiliation{P.N. Lebedev Physical Institute of the Russian Academy of Sciences, Moscow 119991, Russia}
\affiliation{Russian Quantum Center, Skolkovo, Moscow 121205, Russia}

\author{Evgeniy O. Kiktenko}
\affiliation{Russian Quantum Center, Skolkovo, Moscow 121205, Russia}

\author{Ksenia Yu. Khabarova}
\affiliation{P.N. Lebedev Physical Institute of the Russian Academy of Sciences, Moscow 119991, Russia}
\affiliation{Russian Quantum Center, Skolkovo, Moscow 121205, Russia}

\author{Ilya A. Semerikov}
\affiliation{P.N. Lebedev Physical Institute of the Russian Academy of Sciences, Moscow 119991, Russia}
\affiliation{Russian Quantum Center, Skolkovo, Moscow 121205, Russia}

\author{Aleksey K. Fedorov}
\affiliation{P.N. Lebedev Physical Institute of the Russian Academy of Sciences, Moscow 119991, Russia}
\affiliation{Russian Quantum Center, Skolkovo, Moscow 121205, Russia}

\author{Nikolay N. Kolachevsky}
\affiliation{P.N. Lebedev Physical Institute of the Russian Academy of Sciences, Moscow 119991, Russia}
\affiliation{Russian Quantum Center, Skolkovo, Moscow 121205, Russia}

\begin{abstract}
An efficient implementation of the Toffoli gate is of conceptual importance for running various quantum algorithms, including Grover's search and Shor's integer factorization. 
However, direct implementation of the Toffoli gate either entails a prohibitive increase in the number of two-qubit gates or requires ancilla qubits, 
whereas both of these resources are limited in the current generation of noisy intermediate-scale quantum devices. 
Here, we experimentally demonstrate a scalable $N$-qubit Toffoli gate improvement using $^{171}$Yb$^{+}$ trapped-ion-based optical-metastable-ground encoded qutrits for the cases of 
\RC{up to $N$=10.}
With the use of the M$\o$lmer-S$\o$rensen gate as a basic entangling operation, we compare the standard qubit decomposition with the qutrit approach, where upper levels are used as ancillas. 
The presented decomposition requires only global control of the ancilla levels, which simplifies experimental implementation of the proposed approach.
\RC{Using the example of a three-qubit Grover's search, we also demonstrate an increase in the algorithm's accuracy by monitoring the leakage from the qubit subspace during a qutrit-based Toffoli gate implementation.}
\end{abstract}

\maketitle

\RC{{\it Introduction}.
In recent years, significant efforts have focused on building quantum processors using various physical systems to solve computational problems beyond the reach of classical computers.
These systems include superconducting circuits~\cite{Martinis2019,Pan2021}, semiconductor quantum dots~\cite{Loss1998,Vandersypen2022,Morello2022,Tarucha2022}, photonic systems~\cite{Pan2020,Lavoie2022}, neutral atoms~\cite{Lukin2021,Browaeys2021,Browaeys2020-2,Saffman2022}, 
and trapped ions~\cite{Monroe2017,Blatt2012,Blatt2018}.
Despite significant progress, maintaining coherence and precise control while scaling up these systems remains a major obstacle to achieving quantum advantage in practical applications~\cite{Fedorov2022}.
One of the possible approaches to scaling is a transition from a qubit- to qudit-based quantum computing, where quantum information is stored in the multilevel ($d$-level) structure of the aforementioned physical systems (for a review, see Refs.~\cite{Sanders2020,Kiktenko2023}). 
}

\RC{One of the problems that can be addressed with qudits is the resource-efficient decomposition of the generalized $N$-qubit Toffoli gate~\cite{Barenco1995}, which is widely utilized in Grover's search algorithm~\cite{Grover1996}, prime factorization~\cite{Shor1994, Antipov2022}, and implementation of an arbitrary multiqubit gate~\cite{NielsenChuang2000}. 
The use of an auxiliary third level within each information carrier enables one to perform an $N$-qubit Toffoli gate with only $2N-3$ native two-qudit operations and connectivity specified by the physical platform~\cite{Ralph2007, Kiktenko2020, Nikolaeva2022}. 
Recently, such decomposition technique was experimentally realized for up to an $8$-qubit Toffoli gate~\cite{chu2023scalable} with superconducting systems \cite{chu2023scalable, Goss2023Toffoli} and theoretically described for trapped-ion platform \cite{Nikolaeva2023}. }

\RC{A related method for trapped-ion processors, proposed in~\cite{Blatt2009toffoli} and generalized in~\cite{Fang2023}, employs a collective vibrational mode shared by all ions as an ancilla. 
This allowed for the experimental demonstration of the Toffoli gate for $N=5$~\cite{Fang2023}, which was the largest dimension of the Toffoli gate realized in trapped ions so far. It also requires only $2N-3$ two-particle Cirac-Zoller~\cite{CZgate} operations, but also demands an auxiliary multidimensional degree of freedom coupled to all qubits involved. This limits its application in the quantum charged coupled device (QCCD) trapped-ion processor architecture~\cite{kielpinski2002architecture, moses2023race} where sub-arrays of ions are stored in separate traps, and no such common degree of freedom is available.

In this Letter, we report an experimental demonstration of a generalized $N$-qubit Toffoli gate in a trapped-ion quantum processor using qutrits ($d=3$), in which upper levels are used as ancillary states. 
The experiment was conducted on a register of 10 $^{171}\mathrm{Yb}^+$ ions, using optical-metastable-ground (\textit{omg}) qutrit encoding.
In addition to its compatibility with QCCD architecture, the Toffoli gate protocol implemented here utilizes a M$\o$lmer-S$\o$rensen (MS) entangling gate --- native to most trapped-ion processors --- and requires only global control over the ancilla-level, making it readily adoptable by other experiments. We perform such a gate on up to $N=10$ ions and show a scalable improvement in its performance with increase of $N$ in comparison with standard qubit decomposition. We also demonstrate that measuring the population of the ancillary state after applying the gate can be used to further improve its performance. As a basic demonstration, we use the proposed decomposition to implement a three-qubit Grover's search algorithm with three qutrits.
}

{\it Qudit processor.} 
The quantum register is implemented as a string of 10 ytterbium ions in a linear Paul trap. The trap secular frequencies are $\{\omega_x, \omega_y, \omega_z\}=2\pi\times\{3.650, 3.728, 0.129\}$\,MHz. 
Here we use levels

\begin{align*}
    \ket{0} & = \ket{\,^2S_{1/2}(F=0,m_F=0)}, \\
    \ket{1} & = \ket{\,^2D_{3/2}(F=2,m_F=0)} \\
\end{align*}
in each ion, which are coupled via an optical E2 transition at 435.5\,nm, as a qubit (Fig.~\ref{fig:levels}). Each ion can be individually optically addressed \RC{with one of two tightly focused laser beams, enabling both single-qudit and two-qudit operations. 
} An auxiliary level 
\begin{equation*}
\begin{aligned}
    \ket{2} & = \ket{\,^2S_{1/2}(F=1,m_F=0)}
\end{aligned}
\end{equation*}
in each ion is used here as a clean ancilla and extends each ion to an \textit{omg}-qutrit. This level is coupled to the $\ket{0}$ using a global microwave field at 12.6\,GHz.
All of these states have a zero magnetic quantum number and, thus, are insensitive to the magnetic field noise to first order in perturbation theory.

\RC{The coherence of the system is limited by the spontaneous decay from the $\ket{1}$ state with a lifetime of $\tau=53$~ms and the addressing laser phase noise, determining the dephasing time between this level and two others $T_2^*=31(2)$~ms.}

\RC{Before each experimental run, radial modes of ions motion are cooled down to the mean phonon number $\bar{n}<0.1$.
}

The native gate set for our processor consists of single-qudit and two-qudit operations. The first type of single-qudit operations is defined as
\begin{equation}
    R_{\phi}^{0j}(\theta)=\exp(-i\sigma^{0j}_{\phi}\theta/2)
\end{equation}
and can be interpreted as a single-qudit rotation by an angle $\theta$ around an axis, specified by angle $\phi$, in the subspace spanned by levels $\ket{0}$ and $\ket{j}$. 
Here $0$ and $j\in\{1,2\}$ correspond to addressed levels of the qudit, $\sigma_\phi^{0j} = \sigma_x^{0j}\cos\phi  + \sigma_y^{0j} \sin\phi $. The operator 
$\sigma^{0j}_\kappa$ with $\kappa = x,y$ stands for the standard Pauli matrix, which acts within the two-level subspace spanned by $\ket{0}$ and $\ket{j}$ (e.g., $\sigma_y^{0j}=-i\ket{0}\!\bra{j}+i\ket{j}\!\bra{0}$).
For rotations around $x$ and $y$ axes, we fix the following notations:
$R_{x}^{0j}(\theta):= R_{\phi=0}^{0j}(\theta)$, $R_{y}^{0j}(\theta):= R_{\phi=\pi/2}^{0j}(\theta)$. \RC{This type of gate is performed with laser ($R_{\phi}^{01}(\theta)$) or microwave ($R_{\phi}^{02}(\theta)$) pulse resonant to the transition between involved levels. Gate parameters $\phi$ and $\theta$ are controlled with phase and duration of the pulses, respectively. The $R^{0j}_\phi(\pi)$ gate duration  is approximately 10~\textmu s. The fidelities of $R_{\phi}^{01}(\theta)$ and $R_{\phi}^{02}(\theta)$ gates estimated via qubit-like randomized benchmarking are 0.99946(6) and 0.9994(1), respectively.}

In addition to the $R_{\phi}^{0j}(\theta)$ gate, we are also able to implement a virtual phase gate~\cite{McKay2017}, which applies a phase $\theta$ to the qudit state $\ket{j}, j=0,1,2$:
\begin{equation}
    R_{z}^{j}(\theta)=\exp\left(i\theta\ket{j}\bra{j}\right).
\end{equation}
\RC{This gate is performed by shifting phases of all successive laser/microwave pulses.} Since this gate is virtual, its fidelity is assumed to be equal to 1.

A feature of a chosen qutrit encoding is that an ancilla level $\ket{2}$ is coupled to the $\ket{0}$ using a global microwave field. 
Thus, $R_{\phi}^{02}(\theta)$ can be applied only to all ions simultaneously. The virtual {$R_{z}^j(\theta)$ to levels $j=0,2$ can also be applied only to all ions at once for the same reason.

As a two-particle entangling operation, we use the segmented amplitude-modulated M$\o$lmer-S$\o$rensen gate~\cite{sackett2000,Molmer-Sorensen1999,Molmer-Sorensen1999-2,Molmer-Sorensen2000, choi2014optimal} acting on levels $\ket{0}$ and $\ket{1}$:
\begin{multline}
\label{eq:XX_gate_init}
\widetilde{\sf XX}(\chi)=\exp\left[-i\left(\chi\sigma^{01}_x\otimes\sigma^{01}_x+\right.\right.\\
\left.\left.\chi_A (\sigma^{01}_x\cdot\sigma^{01}_x) \otimes\mathbb{I}_3+\chi_B\mathbb{I}_3\otimes(\sigma^{01}_x\cdot\sigma^{01}_x)\right)\right],
\end{multline}
where $\chi_A$ and $\chi_B$ are single-qudit phases acquired only by states $\ket{0}$ and $\ket{1}$ in each ion, and dot denotes standard matrix multiplication \RC{(we note that upon generalization to qudits, the $\sigma^{01}_x\cdot\sigma^{01}_x$ operator ceases to act as the identity operator $\mathbb{I}_3$)}. These additional phases are calibrated and compensated with single-qudit operations [see Supplemental Materials (SM)] on a hardware level, so further in this paper we consider ${\sf XX} (\chi) =\exp(-i\chi\sigma^{01}_x\otimes\sigma^{01}_x)$ to be our native gate. The gate duration is 0.92~ms and qudits' connectivity is full. \RC{The ${\sf XX} (\pi/4)$ gate fidelity was characterized by the fidelity of Bell state preparation. The mean fidelity over a 10 ion register, corrected for state preparation and measurement errors (SPAM), is 96.3(2)\%. The leading contributions to the error are spontaneous decay from $\ket{1}$ state ($\sim 1\%$), addressing laser phase noise ($\sim 1\%$), optical cross-talk ($\sim 0.5\%$) and secular frequencies drifts ($\sim0.5\%$).}

\begin{figure}
\includegraphics[scale=0.5]{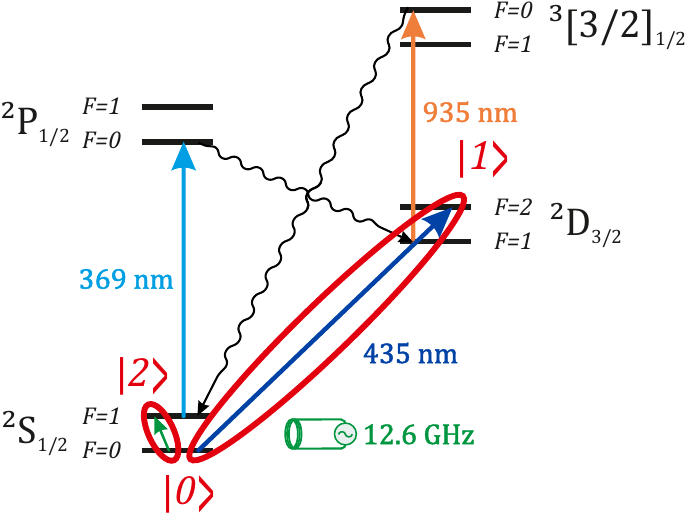}
\caption{The partial level scheme of the $^{171}$Yb$^{+}$ ion. Lasers at 369\,nm and 935\,nm provide Doppler cooling, state initialization, and readout (auxiliary laser sidebands to avoid population trapping in hyperfine components are not shown). Qubit states $\ket{0}$ and $\ket{1}$ are coupled with the laser at 435.5\,nm, while transition between $\ket{0}$ and $\ket{2}$ is driven globally using a microwave antenna. }
\label{fig:levels}
\end{figure}

\RC{At the end of each experiment a standard qubit-like electron shelving readout~\cite{Leibfried2003, semenin2021optimization} is performed on a $^2S_{1/2}\to\,^2P_{1/2}$ transition at 369~nm, as the $\ket{2}$ level should be populated only during the multiqubit gate implementation. Under the illumination of 369~nm and 935~nm laser light, the fluorescence signal from each ion is registered with a multichannel photomultiplier tube (PMT) and ``bright'' ions (strongly scattering photons) are assigned to be measured in state $\ket{0}$, while ``dark'' (non-fluorescing) ions are assigned state $\ket{1}$. However, to study the qutrit gate protocol in more details we also perform a second measurement right after the main one to detect leakage from the qubit subspace. It is implemented by shelving the population from the $\ket{2}$ state to the ``dark'' $^2D_{3/2}(F=2,m_F=-1)$ level before the first readout with a set of single-qudit gates followed by deshelving it back, and a second readout after the main one. Thus, ions being ``dark'' during the first measurement and becoming ``bright'' during the second readout are considered to have leaked from the qubit subspace. Such a protocol was chosen to minimize the influence of the SPAM error on the main readout results, as well as to ensure that the SPAM effect on the leakage estimation would overestimate this value rather than underestimate it. The duration of a single measurement is 0.5~ms. A typical SPAM error is $\sim1\%$ per ion per measurement and is caused by spontaneous decay of the $^2D_{3/2}$ level and non-resonant optical pumping during the process.}
More details about the setup and used methods can be found in the SM and \RC{Ref.~\cite{Zalivako2024, zalivako2025ufn}}. 

{\it Qutrit-based Toffoli gate decomposition.}
The standard three-qubit Toffoli gate ${\sf CCX}$, which inverts the state of the target qubit if both control qubits are in state $\ket{1}$, can be generalized to the $N$-qubit case as follows:
\begin{multline}
    \cnx: \ket{c_1,\ldots,c_{N-1},t}
    \\
    \mapsto \ket{c_1,\ldots,c_{N-1}, t\oplus {\prod}_{i=1}^{N-1} c_i},
\end{multline}
where $c_i,t\in\{0,1\}$ and $\oplus$ stands for modulo 2 addition.
\RC{To decompose this gate, we use a qutrit-based approach. It is scalable, as it can be implemented for an arbitrary $N$, and requires $2N-3$ two-qutrit gates as shown in Fig.~\ref{fig:CNXglobal}(a).
First, we sequentially apply two-qutrit gate $U_1$ to $N-2$ pairs of qutrits. 
Each $U_1$ gate leaves the ``bottom'' qutrit in the state $\ket{1}$ if and only if the qutrits are both in the state $\ket{1}$ prior to $U_1$ operation. 
Thus, after the sequence of $U_1$ gates, the $(N-1)$-th qutrit will be in the state $\ket{1}$ if all the $N-1$ qutrits are initially in the $\ket{1}$ state. 
Then, we apply the $U_2$ gate to the remaining pair of qutrits to invert the state of the $N$-th qutrit whenever $(N-1)$-th qutrit is in the state $\ket{1}$. 
Finally, the uncomputing sequence of $U_1^\dagger$ operations is applied. 
To make the decomposition suitable for trapped ions, it was formulated in terms of ion native gates in \cite{Nikolaeva2023}.
Here we adapt it for a global ancillary levels control by adding a second global $R_\phi^{02}(\pi)$ at the end of each $U_1$ [see Fig.~\ref{fig:CNXglobal}(b)], except the last one, which cancels the effect of $U_1$ on all spectator ions. We also reduce the number of single-qudit gates, leavin
The developed decomposition allows us to realize an $N$-qubit Toffoli gate with $2N-3$ entangling gates. Another technique we use in this work to mitigate spontaneous decay from the $\ket{1}$ state is transferring populations from it to the $\ket{2}$ until they are used in the algorithm (see SM).
Without additional physical systems as ancillas, qubit-based decompositions require $O(N^2)$ gates \cite{Barenco1995}.}

{\it Results.}
To characterize the performance of \RC{$N$-qubit} Toffoli gate realizations and to compare them with qubit decompositions, we \RB{experimentally} reconstruct computational basis truth tables for both qubit and qutrit cases. \RC{
The gate performance is estimated by computing \textit{truth table fidelities}, defined as $\mathcal{F}_{\rm tt}=2^{-N}\sum_{x\in\{0,1\}^N}f_x$, where $f_x$ represents the measured frequency of obtaining the correct output bitstring for an input basis state $\ket{x}$. 
The dependence of $\mathcal{F}_{\rm tt}$ on $N$ is shown in Fig.~\ref{fig:toffoli_scaling}. Qubit decompositions were measured up to $N=6$, where the corresponding $\mathcal{F}_{\rm tt}$ drops to 4.2(1)\%. As $\ket{2}$ is not used during this type of gates, only one measurement at the end of each circuit is performed. 
Truth tables themselves with supporting details are presented at SM.  
All experiments were performed on an array of 10 ions in the trap.} 

\RB{Fig.~\ref{fig:toffoli_scaling} shows that the qutrit-based gate decomposition (red squares) demonstrates a clear advantage in terms of $\mathcal{F}_{\rm tt}$ in comparison to standard qubit variant (blue circles).
The number of two-particle gates in the qutrit-based decomposition ($2N-3$) grows more slowly than in the qubit-based approach, thus $\mathcal{F}_{\rm tt}$ also diminishes more slowly with increasing $N$.
This supports the claim of better scalability of this gate type. 

We also characterize leakage rate and its influence on the $\mathcal{F}_{\rm tt}$.
By leakage we mean the probability that at least one ion leaves the qubit subspace after application of the qutrit Toffoli gate.
Distribution of these probabilities for different initial states of the register, prepared before the gate application, is plotted in Fig.~\ref{fig:leakage_rate}. 
The mean values for the distributions were fitted with an exponential decay function $f(N)=1-A\times p^{2N-3}$, and a value of $p=0.92(1)$ was obtained. Thus, the leakage probability grows exponentially with number of two-qudit operations in the gate.
Possible causes for leakage include a spontaneous decay from the $\ket{1}$ state to either $\ket{0}$ or $\ket{2}$ levels, optical cross-talks, and errors in two-qudit operations.

Since the leakage rate is sensitive to various errors in the system, the presence of leakage can be used as a ``flag'' that an error has occurred.  
The green triangles in Fig.~\ref{fig:toffoli_scaling} show $\mathcal{F}_{\rm tt}$ for qutrit-based Toffoli gate, where only shots {\it without} leakage were post-selected. The improvement of the \textit{truth table fidelity} can be clearly seen in comparison to both raw qutrit and qubit data.}

\newcommand{\ph}[2]{{\sf Ph}^{#1}(#2)}
\newcommand{\xx}[2]{{\sf XX}^{#1}(#2)}
\newcommand{\zz}[2]{{\sf ZZ}^{#1}(#2)}

\newcommand{\gr}[3]{\gate{R_{#1}^{#2}(#3)}}

\newcommand{\grz}[2]{\gate{R_{z}^{#1}(#2\pi)}}
\newcommand{\grx}[2]{\gate{R_{x}^{#1}(#2\pi)}}
\newcommand{\gry}[2]{\gate{R_{y}^{#1}(#2\pi)}}

\tikzset{
operator/.append style={fill=white!20, rounded corners},
}

\begin{figure}
    \centering
\subfloat[]{\resizebox{.8\linewidth}{!}{\begin{quantikz}[wire types={q,q,n,n,q,q,q,q}, align equals at = 4.5, font=\huge, row sep={1cm,between origins}]
&\ctrl[style={scale=2}]{1} &\ghost{U_1^\dagger}\\
&\ctrl[style={scale=2}]{1} &\\
&\vdots    &\\[-0.5ex]
& &\\
&\ctrl[style={scale=2}]{1}\vqw{-1} &\ghost{H_I^I}\\
&\ctrl[style={scale=2}]{1} &\ghost{H_I^I}\\
&\ctrl[style={scale=2}]{1} &\ghost{H_I^I}\\
&\targ[style={scale=2}]{}  &\ghost{H_I^I}
	\end{quantikz}\hspace{3ex}~\begin{huge}{=}\end{huge}~\hspace{3ex}
\begin{quantikz}[wire types={n,n,n,n,n,n,n,n,n},row sep={1cm,between origins}, align equals at = 4.5, font=\huge]
\setwiretype{q}&\gate[wires=2,, style=U1]{U_1}&\qw\setwiretype{n} &\dots\setwiretype{n}&\setwiretype{}&\qw&\qw \setwiretype{q}&\qw&\qw&\qw&\qw\setwiretype{n}&\dots &&\gate[wires=2, style=U1]{U_1^{\dagger}}\setwiretype{q}& \qw\\
\setwiretype{q}&				&\qw \setwiretype{n}&\dots\setwiretype{n} &&\qw 					 &\qw&\qw&\qw &\qw&\qw\setwiretype{n}&\dots &&\setwiretype{q}&\qw\\
&\vdots&&\ddots&&&&\vdots&&&& \reflectbox{$\ddots$}&&\vdots&\\[-1.5ex]
&&&&&\\
\setwiretype{q}	&\qw			& \setwiretype{q}&\setwiretype{n}\dots &&\gate[wires=2,style=U1]{U_1} 	 \setwiretype{q}&&\qw &&\gate[wires=2,style=U1]{ U_1^{\dagger}}& \qw\setwiretype{n}&\dots &&\qw&\qw\\
\setwiretype{q}	&\qw			& \setwiretype{q}&\setwiretype{n}\dots &&\qw&\gate[wires=2,style=U1last]{U_1} 	 \setwiretype{q}&\qw &\gate[wires=2,style=U1lastdag]{ U_1^{\dagger}}& &\qw\setwiretype{n}&\dots &&\qw&\qw \\
&\qw 				&\qw &\dots &&\qw&\setwiretype{q}\qw 					 	 &\gate[wires=2,style=U2]{U_2}&\qw& \qw&\setwiretype{n}\qw&\dots &&\qw&\qw \\
&\qw				&\qw &\dots &&\qw&\qw & \qw                   				& \qw&\qw&\qw&\dots &&\qw&\qw
\end{quantikz}}}\\
\subfloat[]{\resizebox{.72\linewidth}{!}{\begin{quantikz}[wire types = {q,n,q,q,n,q}, row sep={1.5cm,between origins}, font=\huge]\Large
\lstick{}&&\ghost{R^1_y}\\[-0.5cm]
\lstick{}&\vdots&\ghost{R^1_y}\\[-0.3cm]
\lstick{}&\gate[wires=2, style=U1]{U_1}&\ghost{R^1_y}\\
\lstick{}&&\ghost{R^1_y}\\[-0.5cm]
\lstick{}&\vdots&\ghost{R^1_y}\\[-0.3cm]
\lstick{}&&\ghost{R^1_y}
\end{quantikz}
~\begin{huge}{=}\end{huge}~
\begin{quantikz}[wire types = {q,n,q,q,n,q}, row sep={1.5cm,between origins}, font=\huge]
\lstick{}&&\grx{02}{-}\gategroup[6,steps=1,style={rounded
corners,fill=violet!10,draw=violet!20, inner
xsep=2pt},background,label style={label
position=above,anchor=north,yshift=0.8cm}]{{
Global gate}}&&&\grx{02}{}\gategroup[6,steps=1,style={rounded
corners,fill=cyan!10,draw=cyan!20, inner
xsep=2pt},background,label style={label
position=above,anchor=north,yshift=0.8cm}]{{
Global gate}}&\\[-0.5cm]
\lstick{}&&\vdots&&\ghost{R^{10}_y}&\vdots&\\[-0.3cm]
\lstick{}&\gry{01}{-}&\grx{02}{-}&\gry{01}{}&\gate[wires=2]{\xx{}{\frac{\pi}{2}}}&\grx{02}{}&\\
\lstick{}&&\grx{02}{-}&&&\grx{02}{}&\\[-0.5cm]
\lstick{}&&\vdots&&\ghost{R^{10}_y}&\vdots&\\[-0.3cm]
\lstick{}&&\grx{02}{-}&&&\grx{02}{}&
\end{quantikz}
}}\\
\subfloat[]{\resizebox{.98\linewidth}{!}{\begin{quantikz}[wire types = {q,n,q,q,q}, row sep={1.8cm,between origins},font=\huge]
\lstick{}&&\ghost{R^1_y}\\[-0.5cm]
\lstick{}&\vdots&\ghost{R^1_y}\\[-0.3cm]
\lstick{}&&\ghost{R^1_y}\\
\lstick{}&\gate[wires=2, style=U2]{U_2}&\ghost{R^1_y}\\
\lstick{}&&\ghost{R^1_y}
\end{quantikz}
~\begin{huge}{=}\end{huge}~
\begin{quantikz}[wire types = {q,n,q,q,q}, row sep={1.8cm,between origins}, column sep=0.4cm, font=\huge]
\lstick{}&&\grx{02}{}\gategroup[5,steps=1,style={rounded
corners,fill=cyan!10,draw=cyan!20, inner
xsep=2pt},background,label style={label
position=above,anchor=north,yshift=0.8cm}]{{
Global gate}}&&&&&\grx{02}{-}\gategroup[5,steps=1,style={rounded
corners,fill=violet!10,draw=violet!20, inner
xsep=2pt},background,label style={label
position=above,anchor=north,yshift=0.8cm}]{{
Global gate}}&&\\[-0.5cm]
\lstick{}&&\vdots&&\ghost{R^{10}_y}&&&\vdots&&\\[-0.3cm]
\lstick{}&&\grx{02}{}&&&&&\grx{02}{-}&&\\
\lstick{}&\gry{01}{-}&\grx{02}{}&\gry{01}{}&\gate[wires=2]{\xx{}{\frac{\pi}{2}}}&\grx{01}{-}&\gry{01}{-}&\grx{02}{-}&\gry{01}{}&\\
\lstick{}&&\grx{02}{}&&&\grx{01}{-}&&\grx{02}{-}&&
\end{quantikz}}}
    \caption{Qutrit-based decomposition of $\cnx$~gates with $2N-3$ $\sf XX$ gates. All $R_x^{02}(\pm\pi)$ gates are performed globally (depicted with a translucent background). 
    (a) General decomposition scheme.
    In (b) and (c), $U_1$  and $U_2$ are shown, respectively.
    If $N \geq 4$, then in $U_1$ and $U_1^{\dagger}$ the last and the first global gates are not implemented, correspondingly.}
    \label{fig:CNXglobal}
\end{figure}

\begin{figure}
\includegraphics[scale=0.5]{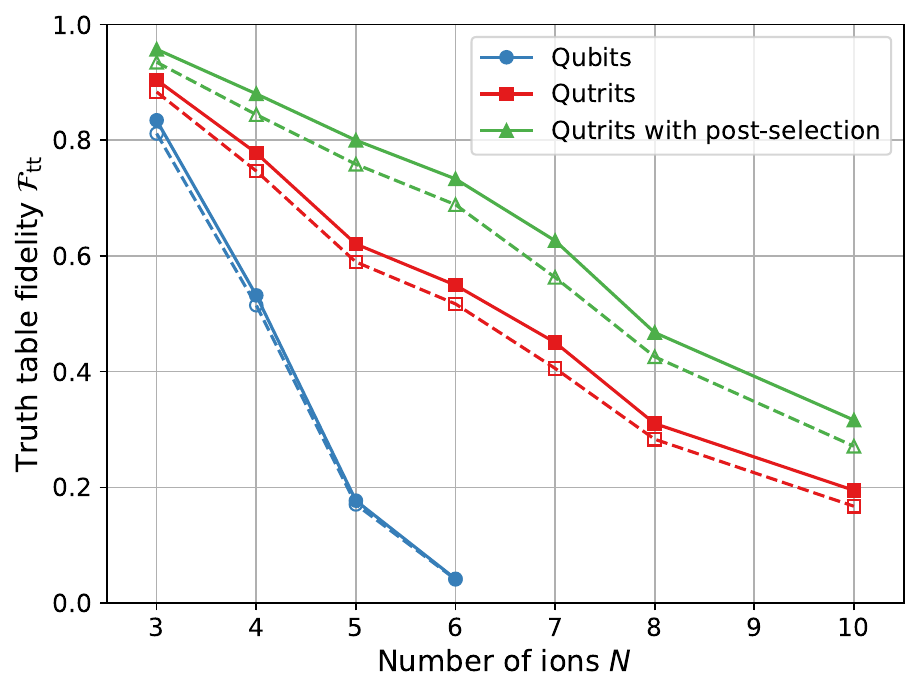}
\caption{\textit{Truth table fidelities} of $N$-qubit Toffoli gates for qubit decomposition (blue circles), qutrit decomposition proposed here (red squares), and qutrit decomposition with postselection (green triangles). Filled markers and solid lines show SPAM-corrected results, empty markers and dashed lines correspond to uncorrected data. The statistical 1-$\sigma$ uncertainties obtained with a bootstrapping method are below 0.3\% for all presented data points.}
\label{fig:toffoli_scaling}
\end{figure}

\begin{figure}
\includegraphics[scale=0.5]{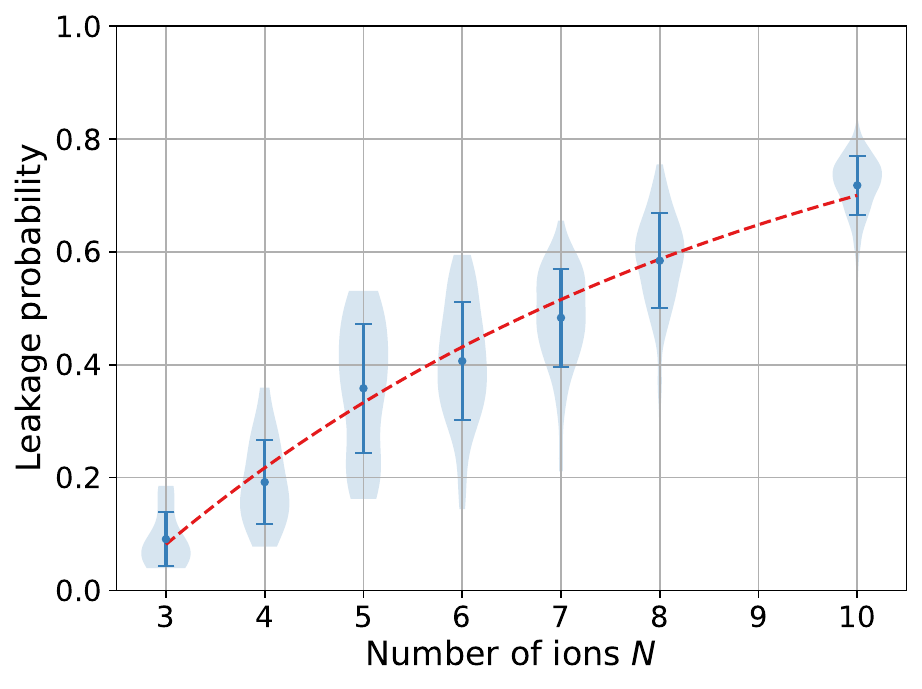}
\caption{The probability of at least one of $N$ ions leakage from the qutrit subspace at the end of the qutrit-based Toffoli gate, as a function of $N$. Distribution of the leakage probability across different initial states of the quantum register is shown for each $N$. Error-bar corresponds for a standard deviation. Red dotted line shows an approximation of the mean leakage probability with function $f(N)=1-A\times p^{2N-3}$, where $p=0.92(1)$ and $2N-3$ --- number of two-qudit gates in the decomposition.}
\label{fig:leakage_rate}
\end{figure}

\renewcommand{\gr}[3]{\gate{R_{#1}^{#2}(#3)}}
\renewcommand{\grz}[2]{\gate{R_{z}^{#1}(\frac{\pi}{#2})}}
\renewcommand{\grx}[2]{\gate{R_{x}^{#1}(\frac{\pi}{#2})}}
\renewcommand{\gry}[2]{\gate{R_{y}^{#1}(\frac{\pi}{#2})}}
\newcommand{\gryy}[2]{\gate{R_{y}^{#1}(-\frac{\pi}{#2})}}
\newcommand{\grxe}[2]{\gate{R_{x}^{#1}(#2)}}
\renewcommand{\xx}[2]{{\sf XX}^{#1}(#2)}
\renewcommand{\zz}[2]{{\sf ZZ}^{#1}(#2)}

\begin{figure*}
\subfloat[]{\resizebox{0.655\linewidth}{!}{
\includegraphics[]{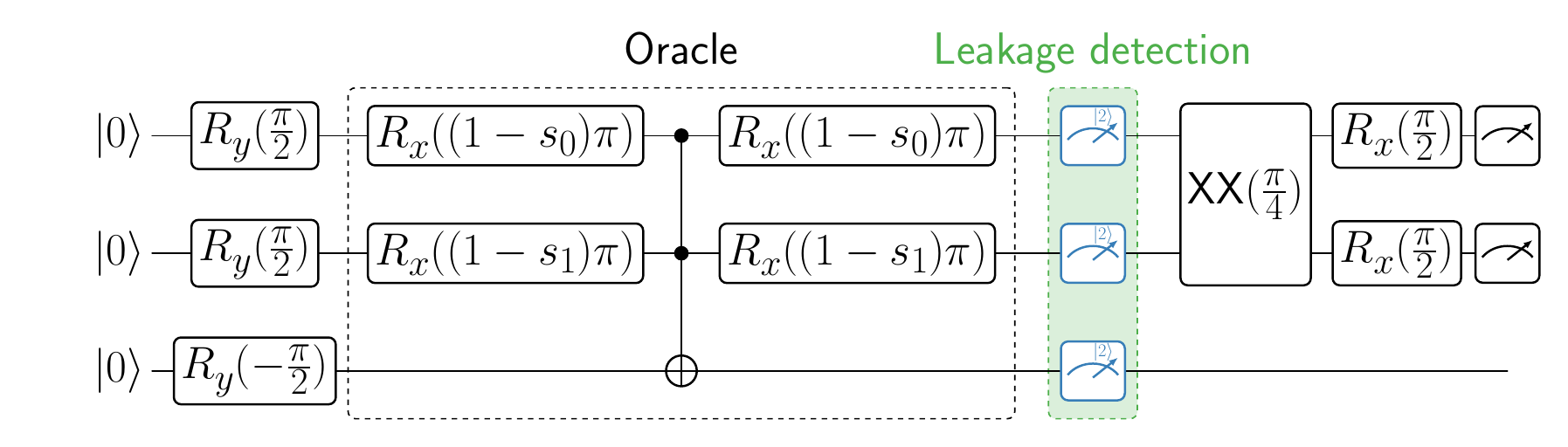}
}}\hfill
\subfloat[]{\resizebox{0.325\linewidth}{!}{
\includegraphics[]{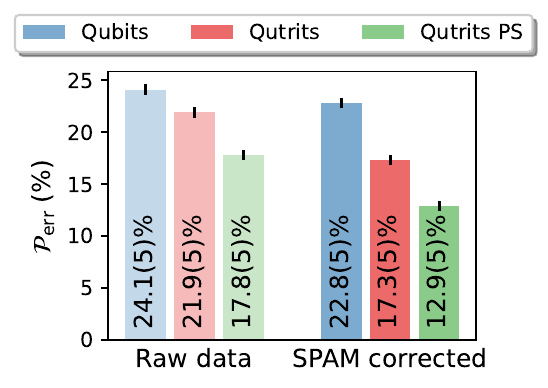}}
}

\caption{(a) Circuit for three-qubit Grover's search algorithm for finding a bitstring $s=(s_0,s_1)$ marked by the oracle with qutrit-based Toffoli gate decomposition. (b) Comparison of Grover's search error ($\mathcal{P}_{\rm err}$) obtained with a qubit-based Toffoli gate decomposition, qutrit-based decomposition and a qutrit-based decomposition with mid-circuit measurement and post-selection.}
\label{fig:Grover}
\end{figure*}

{\it Grover's algorithm implementation.}
\RC{To demonstrate phase properties of qutrit-based Toffoli decompositions, we implement a three-qubit Grover's algorithm with the ${\sf CCX}$ gate in the oracle, which marks a two-bit string $s=(s_0, s_1)$ [see Fig.~\ref{fig:Grover}(a)].
We consider this  implementation as it is highly sensitive to the effect of the Toffoli gate on qubits' phases. 
At the same time, we demonstrate the increased algorithm accuracy after the post-selection procedure.
}

\RC{To implement post-selection, we perform a mid-circuit measurement right after the Toffoli gate. The readout lasers configuration was set in such a way, that only $\ket{2}$ state is ``bright'', thus no heating of the ion chain occurs in the absence of leakage. The Stark shift caused by the readout lasers is compensated via dynamical decoupling (see SM).
}

\RC{The mean error of Grover's algorithm, denoted as $\mathcal{P}_{\rm err}$, for qubit-, qutrit-, and post-selected qutrit-based decompositions of ${\sf CCX}$ is presented in Fig.~\ref{fig:Grover}(b). Each of the four circuits, corresponding to marked bitstrings, was repeated 2048 times.
Qutrit-based decomposition with post-selection provides 1.7 times reduction in the mean search error $\mathcal{P}_{\rm err}$ compared to the qubit-based approach, despite errors introduced by the mid-circuit measurement (blue and green ``SPAM-corrected'' bars). 
The mean fraction of post-selected shots was 0.828(2). 
Only shots with all 10 ``dark'' ions were post-selected to prevent fidelity degradation due to the heating. 
The difference between the Grover's search fidelity in the post-selected qutrit case $1-\mathcal{P}_{\rm err}=87.1(5)\%$ and corresponding Toffoli gate {\it truth table fidelity} $\mathcal{F}_{\rm tt}=95.7(2)\%$ can be explained by additional single- and two-particle gates, imperfections of the mid-circuit measurement, and phase errors in the Toffoli gate. 
This enables us to provide an upper bound on the ${\sf CCX}$ phase error contribution to the algorithm's accuracy of $\sim 5\%$.}

{\it Conclusion.}
We have experimentally implemented a scalable  $N$-qubit Toffoli gate decomposition with up to $N=10$ qutrits, requiring only $2N-3$ entangling operations. 
\RC{The decomposition consists of native ion gates and uses only global control of the ancillary levels, which makes it readily adoptable by other experimental setups and compatible with QCCD architecture.
The direct experimental comparison between qutrit and qubit Toffoli gate decompositions indicates a significant improvement in the gate performance and scalability due to the reduced number of entangling gates. 
For $N > 5$, we were able to obtain the desired patterns of truth tables, which were not possible with qubits in our current setup.}

\RB{
We have also demonstrated how some of the errors that occur during the gate can be tracked by observing the leakage of population from the qubit subspace.
Using post-selection of outcomes without leakage, we obtained a considerable improvement in {\it truth table fidelities} and reduced the mean error in the three-qubit Grover's search.
This demonstrates the method’s applicability to quantum algorithms.
We believe that the development of these methods will pave the way for the efficient utilization of the potential of qudits for implementing quantum algorithms that require many multiqubit operations.
}

{\it Acknowledgements}.
We thank the support of the Russian Roadmap on Quantum Computing
(Contract No. 868-1.3-15/15-2021, October 5, 2021) in the development of the trapped-ion processor. 

\bibliographystyle{apsrev}
\bibliography{bibliography-qudits.bib}

\end{document}


\section*{Supplemental materials}

\subsection{Methods}
\subsubsection{Ions cooling, state initialization and readout} \label{ssec:cooling_readout}
Before each experimental shot, ions are Doppler cooled for 6 ms using a laser beam at 369~nm with a phasemodulation at 14.7~GHz to avoid population trapping in $^2S_{1/2}(F=0)$. Another laser beam at 935~nm phase-modulated at 3.07~GHz provides repumping from the $^2D_{3/2}$ state. After this, ions are initialized into the $\ket{0}$ state by switching off modulation at 14.7~GHz and turning on 2.1~GHz phase modulation of the cooling beam for 20~$\mu s$. This is followed by ground state cooling of all radial modes of motion using the resolved-sideband technique~\cite{Monroe1995} at $\ket{0}\to\ket{1}$ transition to the mean phonon number $\bar{n}<0.1$, and a subsequent another initialization procedure.

At the end of each experimental shot, the output quantum state is retrieved using the electron shelving technique~\cite{Leibfried2003,semenin2021optimization} with a 369~nm light modulated at 14.7~GHz and a repumping beam at 935~nm without modulation for \RC{0.5~ms}. The ions' fluorescence signal at 369~nm is collected with a high numerical aperture lens onto an array of multimode fibers spaced in accordance with the ions' positions in a crystal. Each fiber is coupled to its own channel of the multichannel photomultiplier tube (PMT), connected to the pulse counters. Ions' states are determined by comparison between the registered number of counts and precalibrated threshold values (if the count number is higher than a threshold, the ion is considered to be in the $\ket{0}$ state, and in $\ket{1}$ otherwise). \RC{To measure leakages to the $\ket{2}$ state in experiments with qutrit-based Toffoli gate, its population is transferred to the ``dark'' $\ket{3}=\,^2D_{3/2}(F=2,m_F=-1)$ state before the main readout by a successive application of a global $R_x^{02}(\pi)$ and individual $R_x^{03}(\pi)$ gates to all involved ions. After the first readout, this population is transferred to the ``bright'' $\ket{0}$ state, depopulated during the first measurement, by another series of $R_x^{03}(\pi)$ pulses to all ions. Thus, if an ion was ``dark'' during the first readout and became ``bright'' during the second, it is considered leaked.}

\RC{To perform a mid-circuit detection of the $\ket{2}$ population, a similar procedure is performed, however in this case the modulation at 14.7~GHz of the 369~nm is off. Thus, only state $\ket{2}$ becomes ``bright'' and other states are preserved. It is also important that we discard all shots where fluorescence from any of the ions in the trap was detected. It prevents degradation in the fidelities of the following gates because of the increased temperature. To mitigate the Stark shift occurring due to the readout laser beams, a dynamical decoupling scheme is implemented. For this, the readout process is split into two halves of 0.25~ms each. Between the halves $R_x^{01}(\pi)$ gates are applied to all the ions involved in the experiment. Another series $R_x^{01}(-\pi)$ of gates is applied to the same ions after the second half of the mid-circuit measurement.
More detailed information about the ion trap and experimental procedures can be found in \RC{Ref.~\cite{Zalivako2024, zalivako2025ufn}}. 
}
\subsubsection{Single-qudit operations}
To perform a single-qudit operation $R_{\phi}^{01}(\theta)$, a laser pulse, resonant with the $\ket{0}\to\ket{1}$ transition at 435.5~nm, is applied to a particular ion. Rotation angle $\theta$ is determined by the duration of the pulse, while $\phi$ is set by the pulse carrier phase. Frequency, amplitude, and phase of the laser pulse are controlled using an acousto-optical modulator (AOM). The duration of the $\pi$-pulse is approximately 10~{\textmu}s. The beam waist is 2~{\textmu}m and allows one to individually address a single ion in the chain with a mean cross-talk error below 2\% (ratio between Rabi frequencies on the spectator and target ions). The beam position is controlled with an acousto-optical deflector. The system features two such individual addressing beams to provide both single-qudit and two-qudit operations.

The gate $R_{\phi}^{02}(\theta)$ is performed in a similar way using a global microwave field, resonant with the $\ket{0}\to\ket{2}$ transition at 12.6~GHz. The field is generated using an antenna placed outside of the vacuum chamber. The $\pi$-pulse duration in this case is approximately 10~{\textmu}s.

\RC{The fidelities of $R_{\phi}^{01}(\theta)$ and $R_{\phi}^{02}(\theta)$ gates estimated via qubit-like randomized benchmarking \cite{Siddiqi2021} are 0.99946(6) and 0.9994(1), respectively.}

\RC{In some cases, for example to prepare the initial register states for Toffoli gate truth tables reconstruction, the composite single-qudit gates $R_{\phi}^{01}(\theta)$ implemented with an SK1 technique were used~\cite{brown2004arbitrarily}.}

The virtual gate $R_{z}^{1}(\theta)$ on a particular ion is implemented by shifting the phase of all successive laser pulses involving this particle by the same value. Due to the global addressing of the $\ket{0}\to\ket{2}$ transition, the virtual gates $R_{z}^{0}(\theta)$ and $R_{z}^{2}(\theta)$ can also be performed only on all ions simultaneously, by shifting the phase of the microwave field.

\subsubsection{Two-qudit operations}
As a native two-qudit operation, a M$\o$lmer-S$\o$rensen (MS) gate~\cite{sackett2000,Molmer-Sorensen1999,Molmer-Sorensen1999-2,Molmer-Sorensen2000} on a $\ket{0}\to\ket{1}$ transition is used. It involves the application of bichromatic laser fields to the pair of target ions using two individual addressing beams (one for each particle) orthogonal to the trap axis. The spectral components of the beams are symmetrically detuned from the carrier transition and are close to the red and blue radial secular sidebands. To account for interactions with all motional modes of freedom and reduce sensitivity of the gate fidelity to the parameter drifts, we follow~\cite{choi2014optimal} and amplitude-modulate the pulse in a piecewise constant manner. We use $2N+1=21$ segments with equal duration, where $N=10$ is the number of ions in the trap, resulting in a total gate length of 916~{\textmu}s. Such an operation results in a gate of the form $\widetilde{\sf XX}(\chi)$. \RC{The ${\sf XX} (\pi/4)$ gate fidelity was characterized via Bell state preparation. The mean fidelity over a 10-ion register, corrected for state preparation and measurement errors (SPAM), is found to be 96.3(2)\%. The leading contributions to the error are spontaneous decay from $\ket{1}$ state ($\sim 1\%$), addressing laser phase noise ($\sim 1\%$), optical cross-talk ($\sim 0.5\%$), and secular frequency drifts ($\sim0.5\%$).}

The processor supports gates $\widetilde{\sf XX}(\chi)$ with an arbitrary $\chi$ by scaling the amplitude of the pulse. To perform an $\widetilde{\sf XX}(\chi)$ gate with a negative $\chi$, the carrier phase of the pulse applied to one of the ions is shifted by $\pi$.

To bring the gate $\widetilde{\sf XX}(\chi)$ to the conventional form ${\sf XX} (\chi) =\exp(-\imath\chi\sigma^{01}_x\otimes\sigma^{01}_x)$, the additional phases on levels $\ket{0}$ and $\ket{1}$ were compensated at the hardware level by substituting all ${\sf XX} (\chi)$ gates in the input circuit by the circuits shown in Fig.~\ref{fig:ms-phase-correction}. 

\begin{figure}
	\resizebox{0.6\columnwidth}{!}{
		\begin{quantikz}[font=\huge]
			&\gate[wires=2]{{\sf XX(\chi)}}&\\
			&&
		\end{quantikz}~\begin{huge}=\end{huge}
		\begin{quantikz}[font=\huge]
			&\gate[wires=2]{\sf \widetilde{XX}(\chi)}&\gate{R^{01}_x(\pi)}&\gate{R^{01}_{\chi_A}(-\pi)}&\gate{R^{1}_z(-2\chi_A)}&\\
			&&\gate{R^{01}_x(\pi)}&\gate{R^{01}_{\chi_B}(-\pi)}&\gate{R^{1}_z(-2\chi_B)}&
		\end{quantikz}
	}
	\caption{Correction of the additional phases $\chi_A$ and $\chi_B$, acquired by the states $\ket{0}$ and $\ket{1}$ during two-qutrit MS gate.}
	\label{fig:ms-phase-correction}
\end{figure}

In the case of individual control of the ancilla state $\ket{2}$, this circuit can be further simplified to a pair of virtual gates $R^{2}_z(\chi_A)$ and $R^{2}_z(\chi_B)$.

Calibration of the acquired phases is implemented with a Ramsey-type experiment, given by a circuit in Fig.~\ref{fig:QdPhaseCalibr}. The pair of target qudits is prepared in state $\frac{\ket{0}+\ket{2}}{\sqrt{2}}\ket{2}$, followed by $\widetilde{\sf XX}(\pi/2)$ and an analyzing $\pi/2$-pulse on a $\ket{0}\to\ket{2}$ transition around axis $\phi$, which is scanned. The dependence of the state $\ket{2}$ population on the analyzing pulse phase $\phi$ is fitted by a sinusoidal function, and the phase shift acquired during the two-qudit operation is retrieved. This procedure is repeated for each ion in each two-qudit operation, used in this work.

\newcommand{\gr}[3]{\gate{R_{#1}^{#2}(#3)}}
\newcommand{\grz}[2]{\gate{R_{z}^{#1}(\frac{\pi}{#2})}}
\newcommand{\grx}[2]{\gate{R_{x}^{#1}(\frac{\pi}{#2})}}
\newcommand{\gry}[2]{\gate{R_{y}^{#1}(\frac{\pi}{#2})}}
\newcommand{\gryy}[2]{\gate{R_{y}^{#1}(-\frac{\pi}{#2})}}
\newcommand{\grxe}[2]{\gate{R_{x}^{#1}(#2)}}
\newcommand{\xx}[2]{{\sf XX}^{#1}(#2)}
\newcommand{\zz}[2]{{\sf ZZ}^{#1}(#2)}
\begin{figure}
{\resizebox{.7\linewidth}{!}{
\begin{quantikz}[column sep = 1em,font=\huge]\Large
\lstick{$\ket{0}$}&\grxe{01}{\pi}&\grxe{02}{\frac{\pi}{2}}\gategroup[2,steps=1,style={rounded
corners,fill=teal!10,draw=teal!20, inner
xsep=2pt},background,label style={label
position=above,anchor=north,yshift=0.8cm}]{{
Global gate}}&\grxe{01}{-\pi}&\grxe{02}{\frac{\pi}{2}}\gategroup[2,steps=1,style={rounded
corners,fill=teal!10,draw=teal!20, inner
xsep=2pt},background,label style={label
position=above,anchor=north,yshift=0.8cm}]{{
Global gate}}&\gate[wires=2]{\widetilde{{\sf XX}}(\frac{\pi}{2})}&\gate{R^{02}_{\phi}(-\frac{\pi}{2})}\gategroup[2,steps=1,style={rounded
corners,fill=orange!10,draw=orange!20, inner
xsep=2pt},background,label style={label
position=above,anchor=north,yshift=0.8cm}]{{
Global gate}}&\meter{}\\
\lstick{$\ket{0}$}&&\grxe{02}{\frac{\pi}{2}}&&\grxe{02}{\frac{\pi}{2}}&&\gate{R^{02}_{\phi}(-\frac{\pi}{2})}&
\end{quantikz}
}}
\caption{Circuit for the calibration of the phase acquired by the levels $\ket{0}$ and $\ket{1}$ during the $\widetilde{\sf XX}(\pi/2)$ gate. The part of the circuit before the two-qudit operation prepares state $\frac{\ket{0}+\ket{2}}{\sqrt{2}}\ket{2}$, followed by $\widetilde{\sf XX}(\pi/2)$ and an analyzing $\pi/2$ pulse around axis $\phi$, which is scanned.}
\label{fig:QdPhaseCalibr}
\end{figure}

\subsubsection{SPAM correction}
To correct for State Preparation and Measurement (SPAM) errors \RC{in experiment involving $N$ ions}, we measure a confusion matrix by preparing all \RC{$2^N$} computational basis states followed by computational basis measurement. To reduce the effects of addressing cross-talk and calibration errors, SK1 composite pulses~\cite{brown2004arbitrarily} are applied to prepare the initial states.

The correction matrix is calculated from the confusion matrix by taking its transpose and then inverting it. After that, the output distributions measured in experiments were corrected for SPAM errors by multiplying the correction matrix by the measured distribution vectors.

\subsubsection{Qutrit-based generalized Toffoli gate decomposition}
The basic scheme of the decomposition proposed here is described in Fig.~2 of the main text. Here we present a particular algorithm (Alg.~\ref{alg:qutrits}) used to construct circuits for the generalized Toffoli gate presented in this paper. The blue-colored text in the algorithm is setup-specific and corresponds to an optional feature that we used to partially compensate for the $T_1 = 53$~ms lifetime of the $\ket{1}$ level. This feature can be omitted. This part of the algorithm transfers population from the $\ket{1}$ to $\ket{2}$ during the period when the particular qubits are not used in the decomposition. 

\begin{algorithm}
\caption{Algorithm for building qutrit Toffoli decomposition}\label{alg:qutrits}
\begin{algorithmic}
\Require $N \geq 3$ \Comment{Number of ions}

\Procedure{$U_1$}{$q_1$,$q_2$}
    \State $R_y^{01}$($-\pi$) $\to$ $q_1$ \Comment{{hereinafter, $V\to q_i$ (All) stands for adding gate $V$ to $i$-th (all) qubit(s) of the circuit}}
    \State $R_x^{02}$($-\pi$) $\to$ All
    \State $R_y^{01}$($\pi$) $\to$ $q_1$
    \color{blue}
    \If{$q_1,q_2 = (1,2)$}
        \For{$q = 3, \dots, N$}
            \State $R_y^{01}$($\pi$) $\to$ $q$
        \EndFor
    \Else
        \State $R_y^{01}$($-\pi$) $\to$ $q_2$
    \EndIf
    \color{black}
    \State ${\sf XX} (\frac{\pi}{2}) \to (q_1,q_2)$
    \If{$q_1,q_2 \neq (N-2,N-1)$}
        \State $R_x^{02}$($\pi$) $\to$ All
    \EndIf
\EndProcedure

\Procedure{$U_2$}{$q_1$,$q_2$}
    \color{blue}
    \State $R_y^{01}$($-\pi$) $\to$ $q_2$
    \color{black}
    \State $R_y^{01}$($-\pi$) $\to$ $q_1$
    \State $R_x^{02}$($\pi$) $\to$ All
    \State $R_y^{01}$($\pi$) $\to$ $q_1$
    
    \State ${\sf XX} (\frac{\pi}{2}) \to (q_1,q_2)$
    \State $R_x^{01}$($-\pi$) $\to$ $q_2$
    
    \State $R_x^{01}$($-\pi$) $\to$ $q_1$
    \State $R_y^{01}$($-\pi$) $\to$ $q_1$
    \State $R_x^{02}$($-\pi$) $\to$ All
    \State $R_y^{01}$($\pi$) $\to$ $q_1$
    \color{blue}
    \State $R_x^{01}$($\pi$) $\to$ $q_2$
    \color{black}
\EndProcedure

\For{$q=1,\dots,N-2$}
    \State $U_1(q, q+1)$
\EndFor
\State $U_2(N-1, N)$
\For{$q=N-2,\dots,1$}
    \State ($U_1^\dagger(q, q+1)$)
\EndFor
\end{algorithmic}
\end{algorithm}

\vspace{-0.8cm}
\subsection{Generalized Toffoli gate characterization}

\RC{Here we provide truth tables for the $N$-qubit Toffoli gate with $N = 3,4,5,6,7,8,10$ (see Fig. \ref{fig:c3x}, Fig. \ref{fig:c4x}, Fig. \ref{fig:c7x}) obtained using qubit-, qutrit-based, and qutrit-based decomposition with post-selection.
We summarize the number of shots, truth tables fidelities with and without SPAM correction, and the number of entangling gates for different decompositions in Table \ref{tab:summary}.

To construct qubit truth tables for $N=3,\dots,6$ $\cnx$ gates we used known qubit-based decompositions, taking advantage of the all-to-all connectivity between ions.
The qubit-based decomposition of ${\sf CCX}$ contains 6 two-qubit gates (see Fig. \ref{fig:qubit-ccx}) and the decomposition of ${\sf CCCX}$ contains 14 two-qubit gates \cite{nakanishi2021quantumgate}.
For $N=5$ and $N=6$, we generated qubit-based decompositions using {\sf qiskit} \cite{qiskit2024} and applied the ion-specific transpilation to these circuits. After transpilation for the trapped-ion processor these circuits }contain only $R_{\phi}^{01}(\theta):=R_{\phi}(\theta)$, $R_{z}^{1}(\theta):=R_{z}(\theta)$ and $\widetilde{\sf XX}(\theta)$ gates. 
Note that the action of $\widetilde{\sf XX}(\theta)$ and ${\sf XX}(\theta)$ in the qubit subspace is the same.

\renewcommand{\r}[3]{R_{#1}^{#2}(#3)}

\newcommand{\ph}[2]{{\sf Ph}^{#1}(#2)}
\renewcommand{\xx}[2]{{\sf XX}^{#1}(#2)}
\renewcommand{\zz}[2]{{\sf ZZ}^{#1}(#2)}

\renewcommand{\gr}[3]{\gate{R_{#1}^{#2}(#3)}}

\renewcommand{\grz}[2]{\gate{R_{z}^{#1}(#2)}}
\renewcommand{\grx}[2]{\gate{R_{x}^{#1}(#2)}}
\renewcommand{\gry}[2]{\gate{R_{y}^{#1}(#2)}}

\begin{figure}
    \centering
\resizebox{0.99\columnwidth}{!}{
\begin{quantikz}[font=\huge]
\lstick{ }&\ctrl[style={ scale=2}]{1}&\ghost{R^1_y}\\
\lstick{ }&\ctrl[style={scale=2}]{1}&\ghost{R^1_y}\\
\lstick{ }&\targ[style={scale=2}]{}&\ghost{R^1_y}
\end{quantikz}~~\begin{huge}$=$\end{huge}~\begin{quantikz}[column sep = 0.7em, font=\huge]\Large
\lstick{ }& \gry{}{\frac{\pi}{2}}& \grx{}{\frac{\pi}{4}}&&&&\gate{\xx{}{\frac{\pi}{4}}}&&&&&&\gate{\xx{}{\frac{\pi}{4}}}&\gate[wires=2]{\xx{}{\frac{\pi}{4}}}&&&\gate[wires=2]{\xx{}{\frac{\pi}{4}}}& \gry{}{-\frac{\pi}{2}}&\\
\lstick{ }& \gry{}{\frac{\pi}{2}}&&\gate[wires=2]{\xx{}{\frac{\pi}{4}}}& \grx{}{\frac{\pi}{4}}&&&&&\gate[wires=2]{\xx{}{\frac{\pi}{4}}}& \gry{}{\frac{\pi}{2}}&&&& \grx{}{-\frac{\pi}{2}}& \grz{}{\frac{\pi}{4}}&& \grx{}{\frac{\pi}{2}}&\\
\lstick{ }& \gry{}{\frac{\pi}{2}}&&& \grx{}{-\frac{\pi}{2}}& \grz{}{\frac{\pi}{4}}& \gate{\xx{}{\frac{\pi}{4}}}\wire[u][2]{q}& \grx{}{-\frac{\pi}{2}}& \grz{}{-\frac{\pi}{4}}&& \grx{}{-\frac{\pi}{2}}& \grz{}{\frac{\pi}{4}}&\gate{\xx{}{\frac{\pi}{4}}}\wire[u][2]{q}&& \gry{}{-\frac{\pi}{4}}&\grz{}{-\frac{\pi}{2}}&&&
\end{quantikz}
}
    \caption{Example of ion-specific transpilation of 3-qubit Toffoli gate circuit. }
    \label{fig:qubit-ccx}
\end{figure}

\vspace{-0.3cm}
\subsubsection{Grover's algorithm}

\begin{figure}
\includegraphics[width=0.8\linewidth]{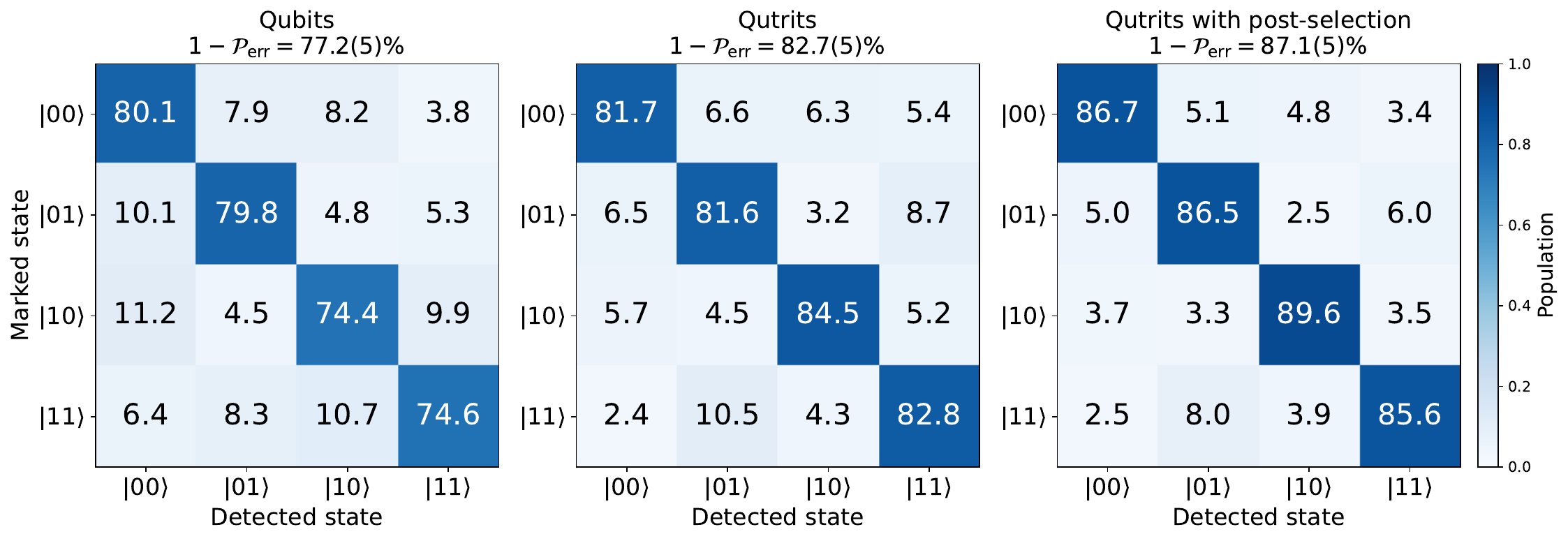}
\caption{Results of the three-qubit Grover's search for different Toffoli gate decompositions. For the qutrit-based Toffoli gate, both cases with and without post-selection are presented. \vspace{-0.1cm}}
\label{fig:Grovel_palette}
\end{figure}

\begin{figure}[h!]
\vspace{-0.2cm}
\resizebox{.99\linewidth}{!}{\includegraphics[]{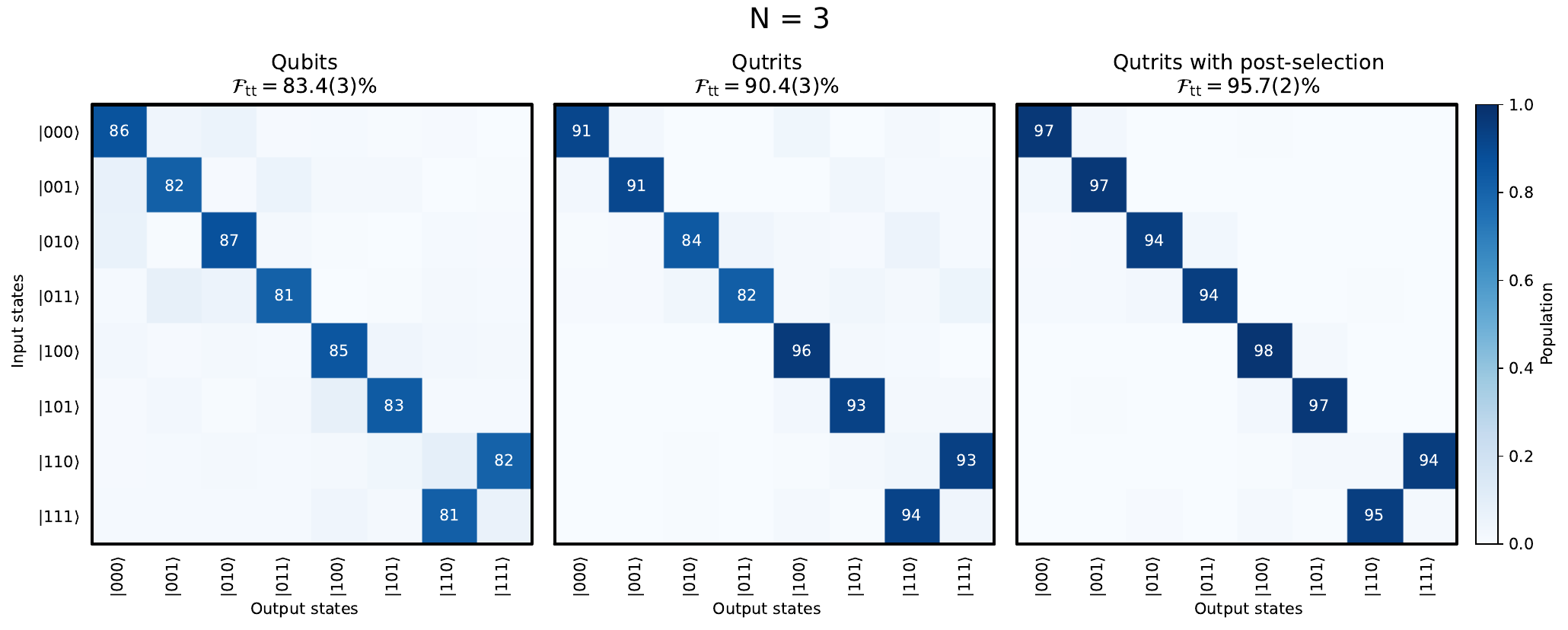}}
\caption{Measured $N$-qubit Toffoli gate truth tables for $N=3$ and various decompositions.}
\label{fig:c3x}
\end{figure}

\begin{figure}
\subfloat[]{\resizebox{.99\linewidth}{!}{\includegraphics[]{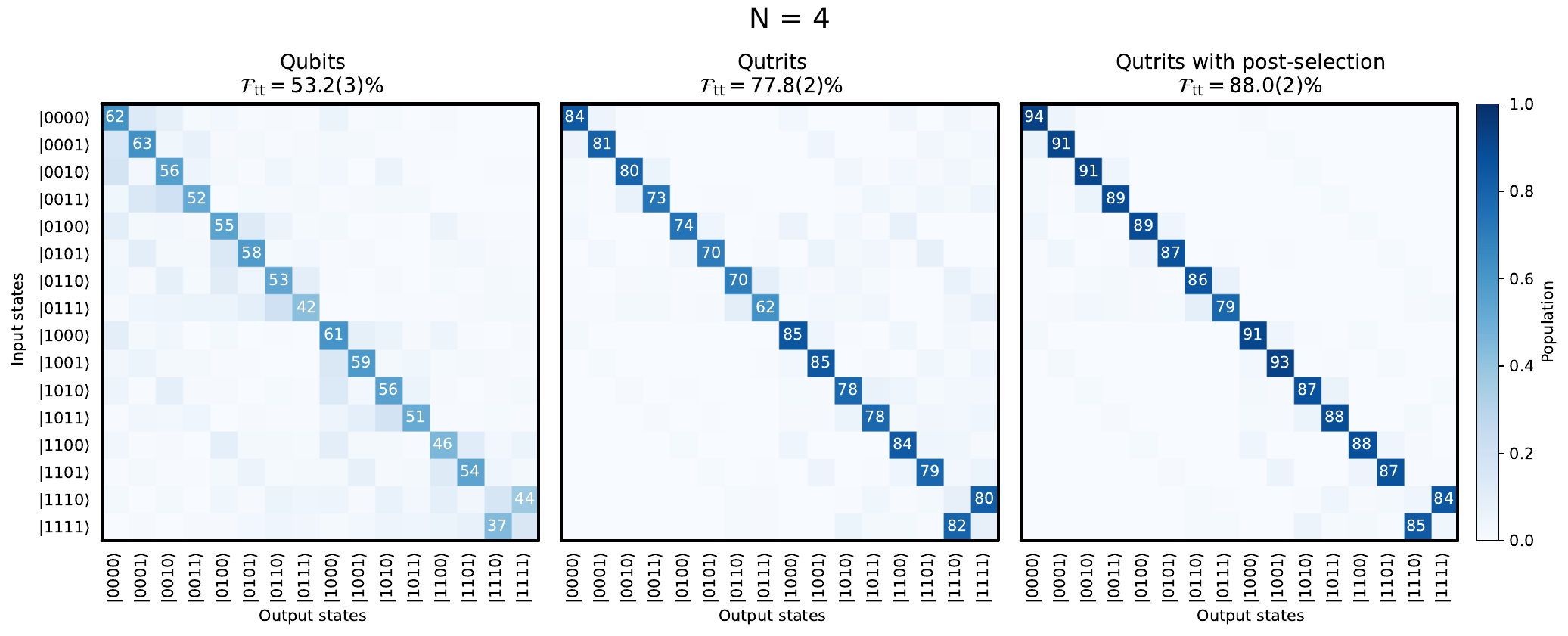}}}\\
\subfloat[]{\resizebox{.99\linewidth}{!}{\includegraphics[]{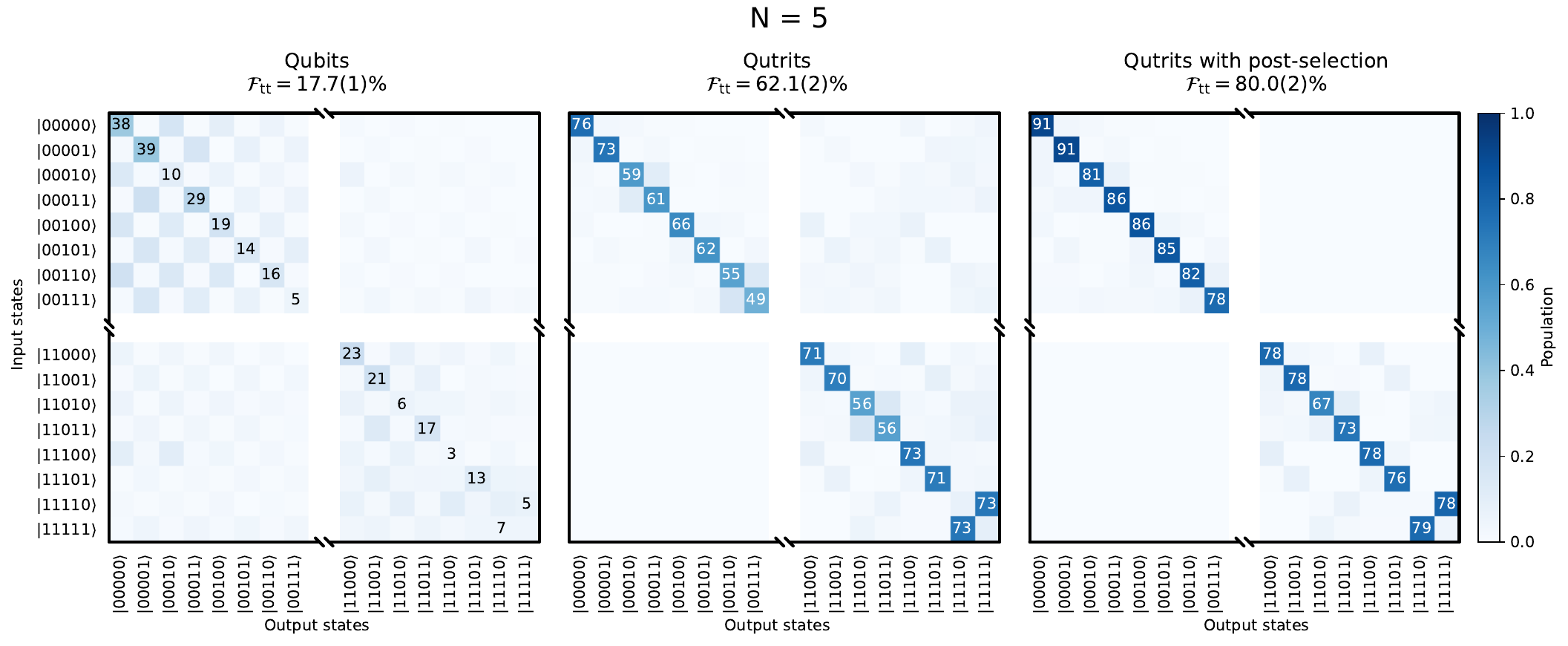}}}\\
\subfloat[]{\resizebox{.99\linewidth}{!}{\includegraphics[]{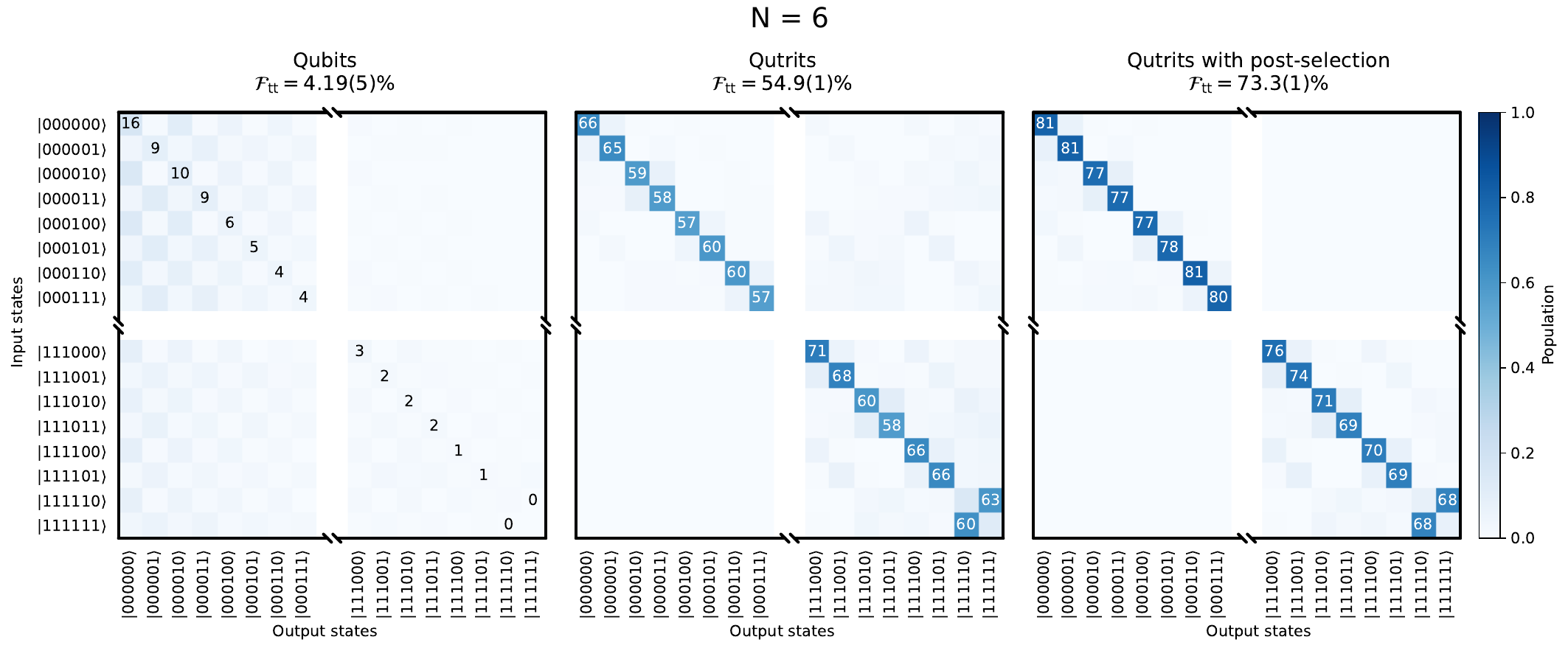}}}
\caption{Measured $N$-qubit Toffoli gate truth tables for $N=4,\dots, 6$ and various decompositions.}
\label{fig:c4x}
\end{figure}

\begin{figure}
\subfloat[]{\resizebox{.66\linewidth}{!}{\includegraphics[]{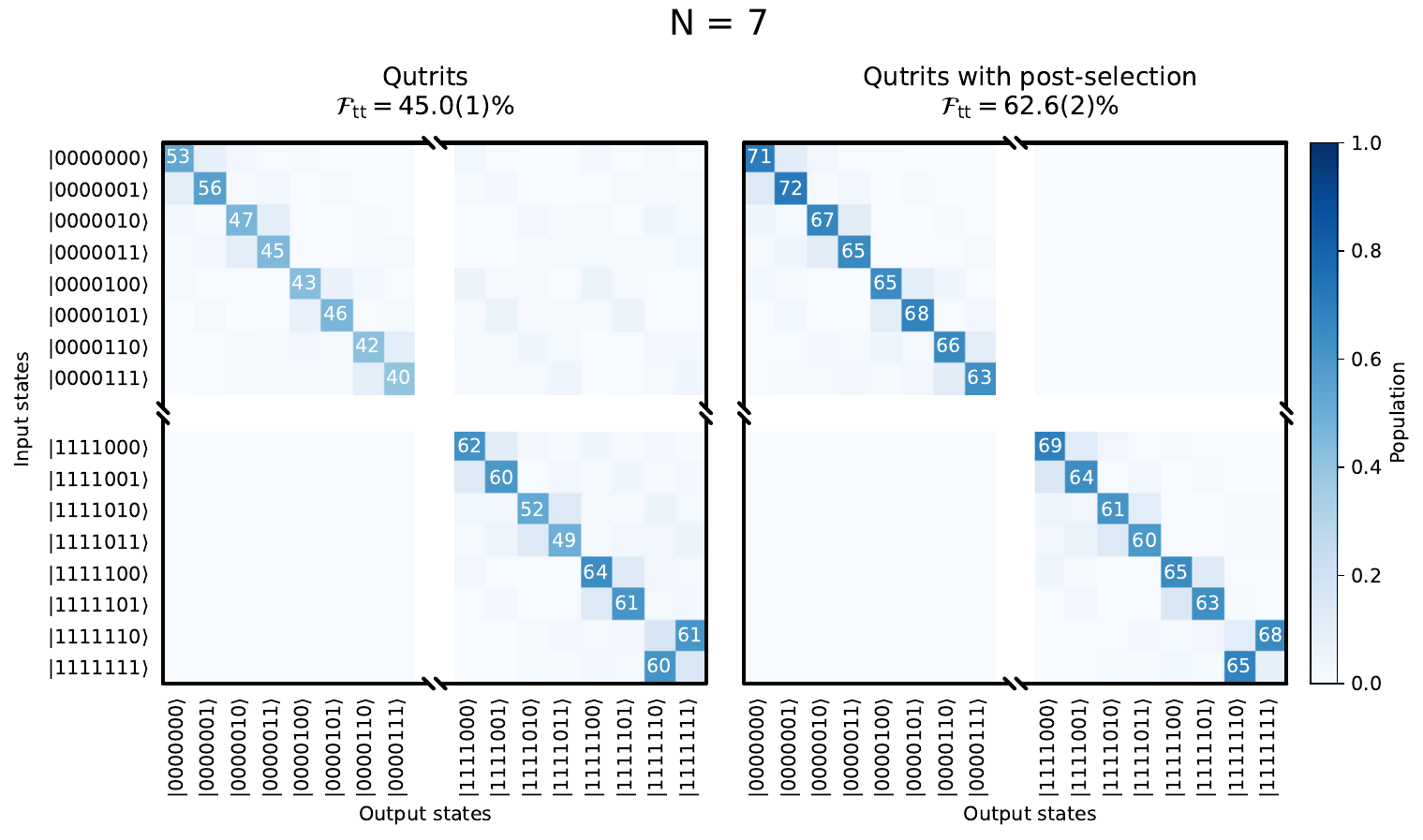}}}\\
\subfloat[]{\resizebox{.66\linewidth}{!}{\includegraphics[]{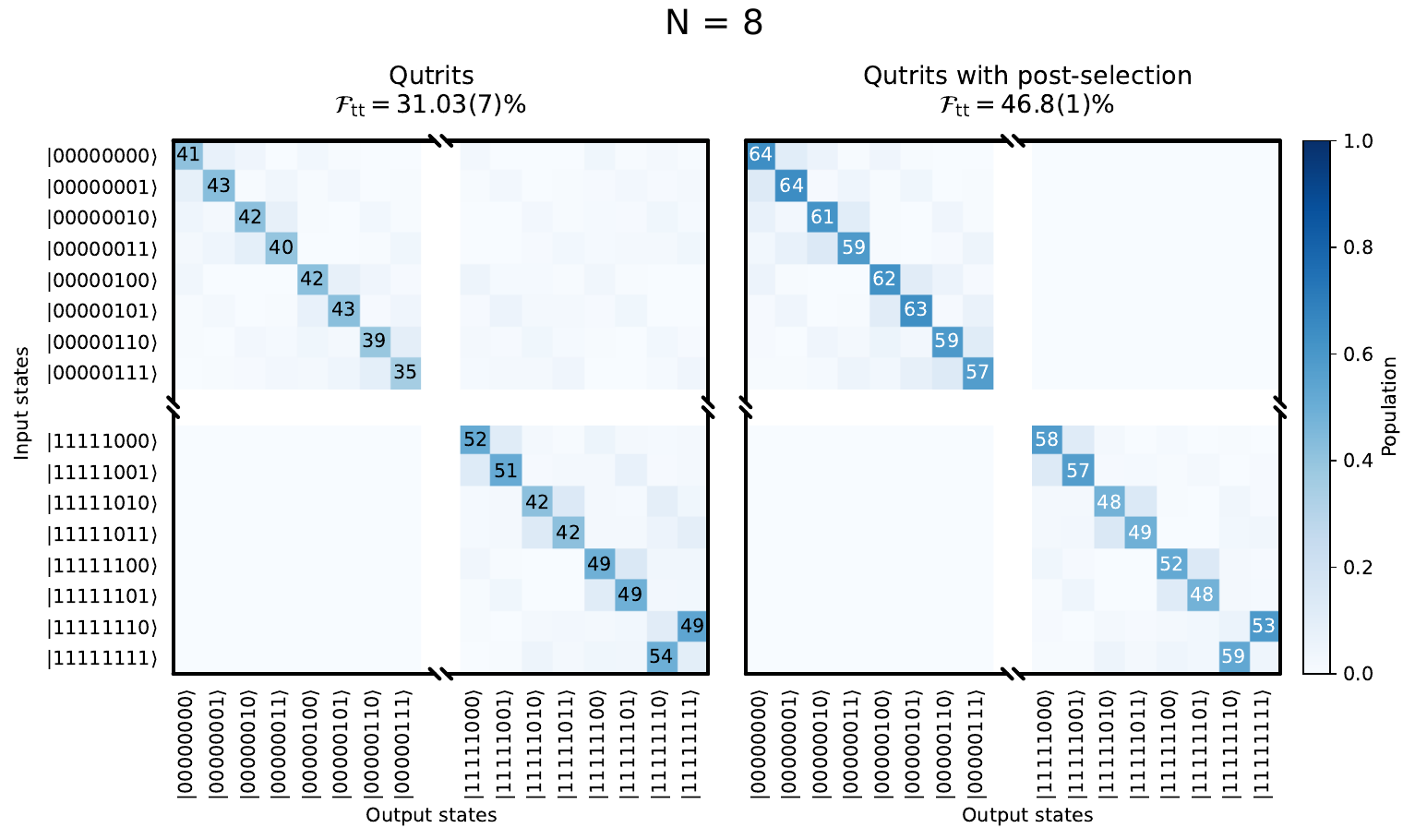}}}\\
\subfloat[]{\resizebox{.66\linewidth}{!}{\includegraphics[]{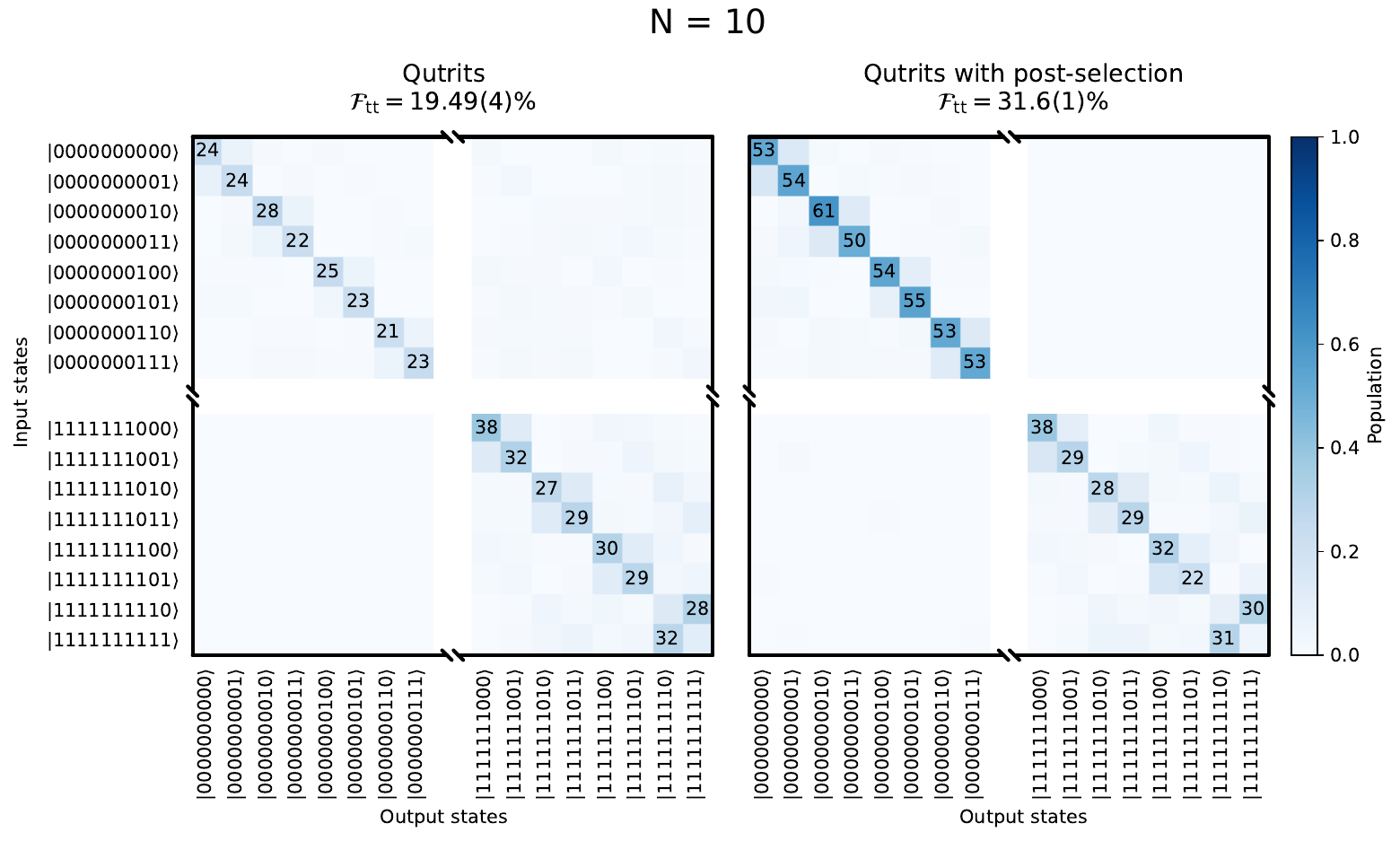}}}
\caption{Measured $N$-qubit Toffoli gate truth tables for $N=7,\dots, 10$ and various decompositions.}
\label{fig:c7x}
\end{figure}

\RC{The SPAM-corrected results of Grover's search algorithm implementation using qubit- and qutrit-based Toffoli gate decomposition as well as using qutrit-based Toffoli gate with post-selection are presented in Fig.~\ref{fig:Grovel_palette}. In each case 2048 shots were performed for each oracle. In the qubit case, decomposition from Fig. \ref{fig:qubit-ccx} was used. In the qutrit-based case without post-selection, only one readout at the end of the algorithm was performed. In the third experiment with post-selection, a mid-circuit measurement was performed after the Toffoli gate, as discussed in Sec.~\ref{ssec:cooling_readout}.} 

\vspace{-0.4cm}

\clearpage

\clearpage
\clearpage

\setlength\tabcolsep{0.13cm}
{\renewcommand{\arraystretch}{1.5}
\begin{table}[]
\begin{tabular}{c|cccc|cccccc|}
\cline{2-11}
                          & \multicolumn{4}{c|}{Qubits}                                                                                                                                    & \multicolumn{6}{c|}{Qutrits}                                                                                                                                                                 \\ \cline{2-11} 
                          & \multicolumn{1}{c|}{\multirow{2}{*}{\shortstack{\!No. of \\ ${\sf XX}(\chi)$ gates\!}}} & \multicolumn{1}{c|}{\multirow{2}{*}{Shots}} & \multicolumn{1}{c|}{\multirow{2}{*}{$\mathcal{F}_{\rm tt}^{\rm raw}$}} & \multirow{2}{*}{$\mathcal{F}_{\rm tt}$} & \multicolumn{1}{c|}{\multirow{2}{*}{\shortstack{\!No. of \\${\sf XX}(\chi)$ gates\!}}} & \multicolumn{1}{c|}{\multirow{2}{*}{Shots}} & \multicolumn{2}{c|}{Without post-selection}                                    & \multicolumn{2}{c|}{With post-selection}    \\ \cline{1-1} \cline{8-11} 
\multicolumn{1}{|c|}{$N$} & \multicolumn{1}{c|}{}                             & \multicolumn{1}{c|}{}                       & \multicolumn{1}{c|}{}                   &                    & \multicolumn{1}{c|}{}                             & \multicolumn{1}{c|}{}                       & \multicolumn{1}{c|}{$\mathcal{F}_{\rm tt}^{\rm raw}$} & \multicolumn{1}{c|}{$\mathcal{F}_{\rm tt}$} & \multicolumn{1}{c|}{$\mathcal{F}_{\rm tt}^{\rm raw}$} & $\mathcal{F}_{\rm tt}$ \\ \hline
\multicolumn{1}{|c|}{3}   & \multicolumn{1}{c|}{6}                             & \multicolumn{1}{c|}{2048}                       & \multicolumn{1}{c|}{81.2(3)\%}                   &        83.4(3)\%            & \multicolumn{1}{c|}{3}                             & \multicolumn{1}{c|}{2048}                       & \multicolumn{1}{c|}{88.3(3)\%}            & \multicolumn{1}{c|}{90.4(3)\%}      & \multicolumn{1}{c|}{93.5(2)\%}  & 95.7(2)\%  \\ \hline
\multicolumn{1}{|c|}{4}   & \multicolumn{1}{c|}{14}                             & \multicolumn{1}{c|}{2048}                       & \multicolumn{1}{c|}{51.5(3)\%}                   &        53.2(3)\%            & \multicolumn{1}{c|}{5}                             & \multicolumn{1}{c|}{2048}                       & \multicolumn{1}{c|}{74.7(2)\%}            & \multicolumn{1}{c|}{77.8(2)\%}      & \multicolumn{1}{c|}{84.4(2)\%}  & 88.0(2)\%  \\ \hline
\multicolumn{1}{|c|}{5}   & \multicolumn{1}{c|}{29}                             & \multicolumn{1}{c|}{2048}                       & \multicolumn{1}{c|}{17.1(1)\%}                   &        17.7(1)\%            & \multicolumn{1}{c|}{7}                             & \multicolumn{1}{c|}{3072}                       & \multicolumn{1}{c|}{59.0(2)\%}            & \multicolumn{1}{c|}{62.1(2)\%}      & \multicolumn{1}{c|}{75.9(2)\%}  & 80.0(2)\%  \\ \hline
\multicolumn{1}{|c|}{6}   & \multicolumn{1}{c|}{61}                             & \multicolumn{1}{c|}{3072}                       & \multicolumn{1}{c|}{4.10(4)\%}                   &        4.19(5)\%            & \multicolumn{1}{c|}{9}                             & \multicolumn{1}{c|}{3072}                       & \multicolumn{1}{c|}{51.7(1)\%}            & \multicolumn{1}{c|}{54.9(1)\%}      & \multicolumn{1}{c|}{68.9(1)\%}  & 73.3(1)\%  \\ \hline
\multicolumn{1}{|c|}{7}   & \multicolumn{1}{c|}{}                             & \multicolumn{1}{c|}{}                       & \multicolumn{1}{c|}{}                   &                    & \multicolumn{1}{c|}{11}                             & \multicolumn{1}{c|}{2048}                       & \multicolumn{1}{c|}{40.58(9)\%}            & \multicolumn{1}{c|}{45.0(1)\%}      & \multicolumn{1}{c|}{56.3(1)\%}  & 62.6(2)\%  \\ \hline
\multicolumn{1}{|c|}{8}   & \multicolumn{1}{c|}{}                             & \multicolumn{1}{c|}{}                       & \multicolumn{1}{c|}{}                   &                    & \multicolumn{1}{c|}{13}                             & \multicolumn{1}{c|}{2048}                       & \multicolumn{1}{c|}{28.32(6)\%}            & \multicolumn{1}{c|}{31.03(7)\%}      & \multicolumn{1}{c|}{42.6(1)\%}  & 46.8(1)\%  \\ \hline
\multicolumn{1}{|c|}{10}   & \multicolumn{1}{c|}{}                             & \multicolumn{1}{c|}{}                       & \multicolumn{1}{c|}{}                   &                    & \multicolumn{1}{c|}{17}                             & \multicolumn{1}{c|}{2048}                       & \multicolumn{1}{c|}{16.74(3)\%}            & \multicolumn{1}{c|}{19.49(4)\%}      & \multicolumn{1}{c|}{27.11(8)\%}  & 31.6(1)\%  \\ \hline
\end{tabular}
\caption{Number of shots and ${\sf XX} (\chi)$ gates in each implemented decompositions and obtained truth table fidelities with ($\mathcal{F}_{\rm tt}$) and without ($\mathcal{F}^{\rm raw}_{\rm tt}$) SPAM-correction.}
\label{tab:summary}
\end{table}

\bibliographystyle{apsrev}
\bibliography{bibliography-qudits.bib}